\documentclass[12pt]{article}
\usepackage[utf8]{inputenc}
\usepackage[english]{babel}
\usepackage{amsthm}
\newtheorem*{remark}{Theorem}

\usepackage{epsfig}
\usepackage[dvipdfmx]{color}
\usepackage{amsmath}
\usepackage{amssymb}
\usepackage{lscape}

\makeatletter
\@addtoreset{equation}{section}

\makeatother

\begin{document}

\title{
\vbox{
\baselineskip 14pt
\hfill \hbox{\normalsize KEK-TH-2017
}} \vskip 1cm
\bf \Large  Secular Terms in Dyson Series to All-Orders of Perturbation
\vskip 0.5cm
}
\author{
Satoshi Iso$^{a,b}$ \thanks{E-mail: \tt iso(at)post.kek.jp}, 
Hikaru Ohta$^{a,b}$ \thanks{E-mail: \tt hohta(at)post.kek.jp}, 
Takao Suyama$^{a}$ \thanks{E-mail: \tt tsuyama(at)post.kek.jp}
\bigskip\\
\it \normalsize
$^a$ Theory Center, High Energy Accelerator Research Organization (KEK), \\
\it  \normalsize 
$^b$Graduate University for Advanced Studies (SOKENDAI),\\
\\
\it Tsukuba, Ibaraki 305-0801, Japan \\
\smallskip
}
\date{\today}

\maketitle

\abstract{\normalsize
In classical and quantum systems,  perturbation of  an evolution equation is often invalidated by
secular terms which diverge at late times. The diverging behavior of evolution can be remedied by
various techniques of resummation such as renormalization group or multi-scale analysis.
In this paper, we prove that, in a  generic quantum mechanical system, 
secular terms can be systematically removed to all orders  in the Dyson  series
by the method of improved (renormalized) perturbation.
A recurrence relation to provide an explicit method to remove the secular terms is given.
As a byproduct, we give a simple method to obtain  energy eigenvalues and decay rates 
to all orders of perturbation.  
}
 
\newpage

\section{Introduction}


The appearance of secular terms is ubiquitous in perturbative calculations \cite{Review1, Review2}.
Secular terms that grow as a power of time $t$ 
would invalidate the use of perturbative series at a later time. 
To remedy this, one usually tries to resum the secular terms so that the resummed series 
is valid at any instant of time. 
The most conventional method is to introduce multiple time scales.
(In the case of quantum anharmonic oscillator, see e.g. \cite{Bender}.) 
A related and  systematic procedure, known as the renormalization group method, 
was proposed in \cite{Chen:1995ena} which can be successfully applied to various classical mechanical systems
and to quantum field theories \cite{deVega}
The appearance of secular terms is also a typical problem in the study of non-equilibrium systems \cite{Berges}. 
In this research area, the resummation is performed by solving the Kadanoff-Baym equations in 
a self-consistent manner. 
The resummation of secular terms is also discussed in the context of quantum field theory on de Sitter space 
\cite{Burgess}.

To illustrate the issues, consider a simple classical anharmonic oscillator 
\begin{equation}
\frac{d^2x}{dt^2}\ =\ -\omega^2x-\lambda x^3. 
   \label{classical AHO}
\end{equation}
For a small $\lambda$, one can solve this equation perturbatively, resulting in 
\begin{equation}
x(t)\ =\ A\sin\omega(t-t_0)+\frac{3A^3}{8\omega}\lambda t\cos\omega(t-t_0)+\cdots. 
   \label{AHO perturbative}
\end{equation}
This has a secular term which grows with $t$ without bound. 
Of course, this is an artifact of the perturbative calculation. 
The equation (\ref{classical AHO}) can be solved explicitly as 
\begin{equation}
x(t)\ =\ C\,{\rm sn}\left[ \frac{\sqrt{2E}}C(t-t_0),k \right], 
\end{equation}
where 
\begin{equation}
C^2\ :=\ \frac{\omega^2}{\lambda}\left[ \sqrt{1+\frac{4E}{\omega^2}\lambda}-1 \right], \hspace{1cm} k^2\ :=\ -\frac {C^4}{4E}\lambda. 
\end{equation}
This exact solution is finite at any $t$. 
One obtains the perturbative series (\ref{AHO perturbative}) by expanding this solution in $\lambda$. 
Therefore, in this case, a reasonable solution is obtained by a resummation of secular terms.

In this paper, we discuss secular terms in quantum mechanics. 
Consider a quantum mechanical system with a Hamiltonian $H_0$ perturbed by a time-independent potential $V$ with a small coupling constant $\lambda$. 
As usual, a straightforward perturbative calculations
generate secular terms in the time evolution of an operator ${ A}(t)$ in the Heisenberg picture
such as 
\begin{equation}
{A}(t) = {A_0}(t) + \sum_{m=1}^{\infty} t^m
\left( \sum_{n=m}^{\infty} \lambda^{n} {A}_{m,n} \right)+ \mbox{(non-secular terms)} .
\end{equation}
where $\lambda$ is a perturbative coupling and $A_0(t)$ is the evolution of $A$ under unperturbed Hamiltonian.
We show that the secular terms which grow in $t$ 
can be resummed systematically to all orders of perturbative expansion 
by appropriately renormalizing the $t$-linear terms $A_{1,n}$ into the unperturbative Hamiltonian.
To achieve this goal, we introduce a modified version of the interaction picture in which the full Hamiltonian is rewritten as 
\begin{equation}
H_0+\lambda V\ =\ H_0(\lambda)+\lambda V(\lambda). 
\end{equation}
The replacement of $H_0$ with $H_0(\lambda)$ amounts to resumming { all} the secular terms  $A_{m,n}$. 
We will show that, by an appropriate choice of $H_0(\lambda)$, we can obtain perturbative expressions for any physical quantities which are free from secular terms. 
In the following, we call an expression without secular terms as  ``secular-free.'' 

There is an interesting by-product of the analysis of resummation. 
The operator $H_0(\lambda)$, which is chosen to eliminate all secular terms, turns out to have eigenvalues which are equal to the energy eigenvalues of this system to all orders in $\lambda$. 
This fact allows us to find an efficient method to calculate the energy eigenvalues of a given perturbed system. 
This method is useful especially when the spectrum of $H_0$ does not have any degeneracies. 
From this formula, it is also straightforward to obtain decaying terms which are higher order generalizations
of Fermi's Golden Rule.

This paper is organized as follows. 
The ordinary perturbation theory is recalled in section \ref{ordinary} to identify the origin of secular terms. 
In section \ref{improved}, we explain our improved perturbation theory. 
This is used in section \ref{all order} to eliminate all secular terms to all orders of the perturbative expansion. 
The section is the main part of the  paper, and we give a recurrence relation in the
improved perturbation to obtain the secular-free evolution equations.
In section \ref{energy}, 
we show that the eigenvalues of $H_0(\lambda)$ give the energy eigenvalues of the system under consideration. 
It is an efficient procedure to calculate the energy eigenvalues to all orders of the  perturbative expansion,  as well as the decay amplitude (i.e. imaginary parts of the energy eigenvalues).
Section \ref{discuss} is devoted to discussion. 
In appendices, we give various detailed calculations. In appendix \ref{Proof-theorem},
we give the proof of the theorem in section \ref{all order}.
In appendix \ref{example}, we study a simple anharmonic oscillator as an example to 
show  the efficiency of our method. 

\vspace{5mm}
\section{Ordinary perturbation theory and secular terms} \label{ordinary}


Let us recall the ordinary perturbation theory in quantum mechanics in order to see how secular terms appear. 

Consider a quantum system with a Hamiltonian of the form  
\begin{equation}
H\ :=\ H_0+\lambda V. 
\end{equation}
We assume that the spectrum of $H_0$ is  discrete  for the time being.
The eigenvalues and eigenstates of $H_0$ are given as 
\begin{equation}
H_0|n\rangle\ =\ E_n|n\rangle. 
\end{equation}
The spectrum may have degeneracies. 
The interaction $V$ is assumed to be time-independent. 

To determine the time evolution of a Heisenberg operator $O_{\rm H}(t)$ perturbatively in $\lambda$, it is convenient to 
employ the interaction picture: 
\begin{equation}
O_{\rm I}(t)\ :=\ U(t)O_{\rm H}(t)U(t)^\dag, \hspace{1cm} U(t)\ :=\ e^{iH_0t}e^{-iHt}. 
\end{equation}
 The operator in the interaction picture
 $O_{\rm I}(t)=e^{iH_0t} O_{\rm I}(0) e^{-iH_0t} $ is assumed to be determined explicitly, which
 does not contain any secular terms. 
Secular terms would appear in the perturbative calculation of the 
unitary operator $U(t)$ which satisfies
 the following differential equation 
\begin{equation}
\frac{d}{dt}U(t)\ =\ -i\lambda V_{\rm I}(t)U(t), \hspace{1cm} V_{\rm I}(t)\ :=\ e^{iH_0t}Ve^{-iH_0t}, 
\end{equation}
with $U(0)=I$, the identity operator. 
This equation can be solved perturbatively as  the Dyson series;
\begin{equation}
U(t)\ =\ I+(-i\lambda)\int_0^tdt_1\,V_{\rm I}(t_1)+(-i\lambda)^2\int_0^tdt_1\int_0^{t_1}dt_2\,V_{\rm I}(t_1)V_{\rm I}(t_2)+O(\lambda^3). 
   \label{ordinary series}
\end{equation}

Secular terms in the operator $O_{\rm H}(t)$ thus come from the integrations in 
the Dyson series of the operator $U(t)$. 
To see the origin of secular terms in $U(t)$ more explicitly, we decompose the potential $V$ as 
\begin{equation}
V\ =\ \sum_{a} V_a, \hspace{1cm} [H_0,V_a]\ =\ \omega_aV_a, 
   \label{decomposition}
\end{equation}
where $\omega_a$ is a difference of eigenvalues of $H_0$, and 
\begin{equation}
V_a\ =\ \sum_{E_m-E_n=\omega_a}|m\rangle\langle m|V|n\rangle\langle n|. 
\end{equation}
The summation over $a$ corresponds to summing over all values of the difference of pairs of eigenvalues of $H_0$. 
In particular we set $\omega_{a}=0$ for $a=0$.  
Thus all pairs of the form $(E_n,E_n)$ are included in $V_0.$ 
Furthermore we denote  $\omega_{-a}=- \omega_a$.

Using the decomposition, $V_I(t)$ is given by
\begin{equation}
V_I(t) \ =\ \sum_{a} V_a e^{i \omega_a t}  \ ,
\end{equation}
the second term in (\ref{ordinary series}) can be written as 
\begin{equation}
-i\lambda\int_0^tdt_1\,V_{\rm I}(t_1)\ =\ -\lambda\sum_{a\ne0}\frac1{\omega_a}\left( e^{i\omega_at}-1 \right)V_a-i\lambda\, tV_0. 
\end{equation}
This shows that a secular term proportional to $\lambda t$ appears from $V_0$ in the first order of perturbation. 

Similarly, secular terms proportional to $(\lambda t)^2$ and also to $\lambda^2 t$
appear from the third term in (\ref{ordinary series}), i.e. the second order perturbation.
From the structure of the Dyson series, 
 we see that all the secular terms 
can be resummed as an exponentiated form $e^{-i f(\lambda) t}$, namely in the
second order of perturbation, the $(\lambda t)^2$ term should be cancelled if 
the $\lambda t$ term is appropriately treated (by renormalization) at the first order of perturbation. 

Thus the structure is similar to the leading-log  resummation of  renormalization in quantum field theories.
The coefficient of the $\lambda t$ terms  corresponds to the  beta-function at the leading order,
and the $\lambda^2 t$ terms correspond to the next-to-leading order terms. 
In general, higher-order secular terms $\lambda^n t^m$ with $n \ge m$
appear from multi-dimensional integrals over $t$, but the only relevant secular terms are
those with $\lambda^n t$: all other others are automatically cancelled. 

\vspace{5mm}
\section{Improved perturbation theory} \label{improved}

In this section, we introduce a modification of the interaction picture which will be used in the next section to eliminate secular terms to all orders in $\lambda$. 

The modification is based on the fact that there is a freedom in decomposing the full Hamiltonian $H$ into a ``free'' part and an ``interaction'' part. 
We employ the following decomposition: 
\begin{equation}
H\ =\ H_0(\lambda)+\lambda V(\lambda), 
\end{equation}
where 
\begin{eqnarray}
H_0(\lambda) 
&:=& H_0+\sum_{n=1}^\infty \lambda^nH_n(\lambda), \nonumber \\
 \lambda V(\lambda) 
&:=&  \lambda(V -H_1(\lambda))
-\sum_{n=2}^\infty\lambda^{n} H_{n}(\lambda).
   \label{decomposition ansats}
\end{eqnarray}
Note that we allow $H_n(\lambda)$ to depend on $\lambda$. 
The $\lambda$-dependence of $H_n$ is irrelevant in the discussion of this section, but 
 becomes  necessary in later sections since
the energy eigenvalue of the improved Hamiltonian $H_0(\lambda)$ depends on $\lambda$ itself. 
 (See e.g. Eqs.(\ref{lastresults})). 
In the next section, we will show how we can systematically determine $H_n(\lambda)$ to all orders
of perturbation so as to eliminate all the secular terms. 

According to this decomposition, we define a modified version of the interaction picture in which operators are evolved by $H_0(\lambda)$ as 
\begin{equation}
O(\lambda,t)\ :=\ e^{iH_0(\lambda)t}O(0)e^{-iH_0(\lambda)t}. 
\end{equation}
As long as the eigenvalues of $H_0(\lambda)$ are known to any desired order in $\lambda$, the right-hand side is given as a sum of exponentials of the form $e^{i\omega t}$ where $\omega$ is the difference of a pair of eigenvalues of $H_0(\lambda)$. 
In other words, $O(\lambda,t)$ is secular-free at this order of perturbation. 
The operator $O(\lambda,t)$ is related to the Heisenberg operator $O_{\rm H}(t)$ as 
\begin{equation}
O_{\rm H}(t)\ =\ U(\lambda,t)^\dag O(\lambda,t)U(\lambda,t), \hspace{1cm} U(\lambda,t)\ :=\ e^{iH_0(\lambda)t}e^{-iHt}. 
\end{equation}
If $U(\lambda,t)$ is secular-free, then so is $O_{\rm H}(t)$. 
Our task is then to show that $U(\lambda,t)$ is secular-free. 

%


The unitary operator $U(\lambda,t)$ is the solution of the following differential equation: 
\begin{equation}
\frac{d}{dt}{U(\lambda,t)}\ =\ -i\lambda V(\lambda,t)U(\lambda,t), \hspace{1cm} V(\lambda,t)\ :=\ e^{iH_0(\lambda)t}V(\lambda)e^{-iH_0(\lambda)t} 
   \label{modified Dyson}
\end{equation}
with $U(\lambda,0)=I$. 
This equation can be solved perturbatively as 
\begin{eqnarray}
U(\lambda,t) 
&=& I+(-i\lambda)\int_0^tdt_1\,V(\lambda,t_1)+(-i\lambda)^2\int_0^tdt_1\int_0^{t_1}dt_2\,V(\lambda,t_1)V(\lambda,t_2) \nonumber \\
&+&(-i\lambda)^3\int_0^tdt_1\int_0^{t_1}dt_2\int_0^{t_2}dt_3\,V(\lambda,t_1)V(\lambda,t_2)V(\lambda,t_3)
+O(\lambda^4). \nonumber \\
\label{improvedDyson}
\end{eqnarray}
This is similar to the ordinary Dyson series (\ref{ordinary series}), but $V(\lambda, t)$ itself 
is $\lambda$-dependent as in the renormalized perturbation. 

\vspace{5mm}

Note that if one expands $H_0(\lambda)$ in $V(\lambda,t)$ as a series of $\lambda$ completely, 
then the factor $e^{iH_0(\lambda)t}$ produces lots of secular terms. 
This factor instead should be kept intact since it is regarded as a result of resummation of secular terms. 
Therefore, we should organize the terms in $U(\lambda,t)$ in a different manner. 

In order to perform the improved perturbation, we introduce the following notations.
First, we introduce an auxiliary variable $\xi$ and deform $\lambda V(\lambda)$ as 
\begin{equation}
\lambda V(\lambda;\xi)\ :=\ (\xi \lambda) \left[ V 
-\sum_{n=0}^\infty (\xi \lambda)^nH_{n+1}(\lambda) \right]. 
   \label{modified V}
\end{equation}
We replace $V(\lambda)$ in $U(\lambda,t)$ with $V(\lambda;\xi)$,
and denote the resulting operator as $U(\lambda,t;\xi)$. 
Obviously, $U(\lambda,t;1)=U(\lambda,t)$ holds. 
Then we define $U_n(\lambda, t)$ by the expansion,
\begin{equation}
U(\lambda,t;\xi)\ =\ I+\sum_{n=1}^\infty (-i \xi \lambda)^nU_n(\lambda,t). 
   \label{xi-expansion}
\end{equation}
The introduction of $\xi$ is simply to count the orders of 
perturbation\footnote{ A naive expansion with respect to $\lambda$ cannot count 
the orders of perturbation because, as we see in Appendix \ref{higher orders}, the definition of
$H_n$ itself contains $\lambda$-dependence through renormalized energy eigenvalues. 
The parameter $\xi$ is introduced to distinguish these two different sources of $\lambda$-dependence.}.

Some examples of the operators $U_n(\lambda,t)$ for small $n$ defined in (\ref{xi-expansion}) are given as 
\begin{eqnarray}
U_1(\lambda,t) 
&&= \int_0^tdt_1\,e^{iH_0(\lambda)t_1}(V-H_1(\lambda))e^{-iH_0(\lambda)t_1},  \\
U_2(\lambda,t) 
&&= \int_0^tdt_1\,e^{iH_0(\lambda)t_1}(-iH_2(\lambda))e^{-iH_0(\lambda)t_1} \nonumber \\
& & +\int_0^tdt_1\int_0^{t_1}dt_2\,e^{iH_0(\lambda)t_1}(V-H_1(\lambda))e^{-iH_0(\lambda)t_1} \nonumber \\
&&\hspace{3cm}\times e^{iH_0(\lambda)t_2}(V-H_1(\lambda))e^{-iH_0(\lambda)t_2},
\end{eqnarray}
\begin{eqnarray}
U_3(\lambda,t) 
&&= \int_0^tdt_1\,e^{iH_0(\lambda)t_1}H_3(\lambda)e^{-iH_0(\lambda)t_1} \nonumber \\
& & +\int_0^tdt_1\int_0^{t_1}dt_2\,e^{iH_0(\lambda)t_1}(-iH_2(\lambda))e^{-iH_0(\lambda)t_1} \nonumber \\
&&\hspace{3cm}\times e^{iH_0(\lambda)t_2}(V-H_1(\lambda))e^{-iH_0(\lambda)t_2} 
\nonumber \\
\label{U1U2U3}
& &
 +\int_0^tdt_1\int_0^{t_1}dt_2\int_0^{t_2}dt_3\,e^{iH_0(\lambda)t_1}(V-H_1(\lambda))e^{-iH_0(\lambda)t_1} 
 \nonumber \\
& & \times
 e^{iH_0(\lambda)t_2}(V-H_1(\lambda))e^{-iH_0(\lambda)t_2} 
 e^{iH_0(\lambda)t_3}(V-H_1(\lambda))e^{-iH_0(\lambda)t_3},\nonumber \\ 
   \label{improved series}
\end{eqnarray}
and so on. 
This corresponds to keeping $H_0(\lambda)$ intact in perturbative expansions.
To show that $U(\lambda,t)$ is secular-free, it is sufficient to show that each $U_n(\lambda,t)$ is secular-free.

This reorganization of the perturbative series has the following advantage. 
Naively, as in the case of $U(t)$ mentioned at the end of section \ref{ordinary}, the multiple integration over $(t_1, t_2, \cdots, t_n)$ in $U_n(\lambda, t)$ 
may possibly generate various types of secular terms proportional to $t^m$ where $m \le n.$ However, we can show that
the multiple integration in $U_n(\lambda,t)$ generates only secular terms proportion to a single power
of $t$ and all the other higher-power terms, $t^m$ with $m \ge 2$, are 
automatically cancelled once the secular terms in $U_l(\lambda,t)$ for $l <n$
are appropriately removed by $H_l(\lambda)$. 

It can be seen as follows.
The second term in $U_2(\lambda,t)$ can be written as
\begin{eqnarray}
\int_0^tdt_1\,
e^{iH_0(\lambda)t_1}(V-H_1(\lambda))e^{-iH_0(\lambda)t_1} U_1(\lambda,t_1) .
\label{U2secular}
\end{eqnarray}
But, if $H_1(\lambda)$ is appropriately chosen, 
$U_1(\lambda, t_1)$ does not generate a secular term and  the integral of (\ref{U2secular}) 
at most generates a term linear in $t$. Similarly, the second term, and a sum of the third and the fourth terms 
in $U_3(\lambda, t)$ are written respectively as
\begin{eqnarray}
&& \int_0^t dt_1\,
e^{iH_0(\lambda)t_1}( -iH_2(\lambda) )e^{-iH_0(\lambda)t_1} U_1(\lambda,t_1), 
\label{U3secular1}
\\
&&
\int_0^tdt_1\,
e^{iH_0(\lambda)t_1}(V-H_1(\lambda))e^{-iH_0(\lambda)t_1} U_2(\lambda,t_1) .
\label{U3secular2}
\end{eqnarray}
Therefore, if both of $U_1(\lambda, t)$ and $U_2(\lambda, t)$ do not generate secular terms,
the multiple integration of $U_3(\lambda, t)$ is reduced to a single integration over $t_1$.

The same argument can be applied to show that each term $U_n(\lambda,t)$ at most generates
only secular terms proportional to a single power of $t$.
Indeed, all terms in $U_n(\lambda,t)$ can be written in terms of a single integration like the above ones. 
 { In the next section} we show this property 
by solving differential equations for $U_n(\lambda,t)$.

\vspace{5mm}
\section{Resummation of secular terms to all orders} \label{all order}

Let us now look at details of each $U_n(\lambda,t)$ to determine explicit forms of $H_n(\lambda)$.
In the following, we will observe that the absence of secular terms alone does not determine $H_n(\lambda)$ uniquely \cite{Pathak}. 
This ambiguity can be fixed by requiring 
\begin{equation}
[H_0, H_0(\lambda)]\ =\ 0
   \label{commutativity}
\end{equation}
to all orders in $\lambda$, namely $[H_0, H_n(\lambda)]=0$ for all $n\ge1$. 
 This commutativity condition is important for a systematic proof of
the absence of secular terms, and also useful for explicit calculations.

Let us briefly summarize our strategy for constructing $H_n(\lambda)$ to all orders in $\lambda$. 
First, we show that $H_n(0)$ can be chosen appropriately such that $U_n(0,t)$ are secular-free for $n\ge1$. 
{This corresponds to taking $\lambda \rightarrow 0$ limit with $\xi \lambda$ in (\ref{modified V})  fixed,
and thus corresponds to replacing $H_0 (\lambda)$ by $H_0 (0)$ in (\ref{U1U2U3})}. 
This is a necessary condition for $U_n(\lambda,t)$ to be secular-free. 
Then, we show that this is also a sufficient condition, 
and that $H_n(\lambda)$ can be obtained from $H_n(0)$ via a simple procedure:
{ namely renormalization of the energy eigenvalues.} 
{In this way we can show that 
 all secular terms are eliminated in our improved perturbation theory. }

 
\subsection{Explicit calculations of $H_1(0)$ and $H_2(0)$}
First we evaluate the multiple $t$-integrations explicitly and  show that $U_1(0,t)$ and $U_2(0,t)$
are secular-free, provided that appropriate $H_1(0)$ and $H_2(0)$ are chosen. 
Similar but more complicated calculations for derivations of $H_3$ and $H_4$ 
are given in Appendices \ref{H3integration} and \ref{4th}. 

The operator $U_1(0,t)$ is given as 
\begin{equation}
U_1(0,t)\ =\ \int_0^tdt_1\,e^{iH_0t_1}(V-H_1(0))e^{-iH_0t_1}. 
\end{equation}
Using the decomposition (\ref{decomposition}), the right-hand side can be written as 
\begin{equation}
\sum_{a\ne0}\frac1{i\omega_a}\left( e^{i\omega_at}-1 \right)V_a+tV_0-\int_0^tdt_1\,e^{iH_0t_1}H_1(0)e^{-iH_0t_1}. 
   \label{U_1 1st}
\end{equation}
The second term can be canceled by choosing 
\begin{equation}
H_1(0)\ =\ V_0. 
\label{H1-0th}
\end{equation}
{Recall that, if there is no degeneracy of energy eigenstates,
 $V_0$ is  a diagonal part of $V$ and given by $V_0=\sum_m   \langle m |V|m \rangle \  |m \rangle \langle m|  .$}
Since $V_0$ commutes with $H_0$, the commutativity condition (\ref{commutativity}) is satisfied for this choice,
up to order ${\cal O}(\lambda)$. 
Note that one may add an operator ${\hat O}$ to $H_1(0)$ without producing extra secular terms 
if and only if ${\hat O}$ does  not commute with $H_0$. 
This ambiguity is fixed by the commutativity condition (\ref{commutativity}). 


Next, we examine $U_2(0,t)$.  This can be written as 
\begin{eqnarray}
U_2(0,t) 
&=& \int_0^tdt_1\,e^{iH_0t_1}(-iH_2(0))e^{-iH_0t_1} \nonumber \\
 &+& \int_0^t  dt_1 \int_0^{t_1}dt_2\,e^{iH_0t_1}(V-H_1(0))e^{-iH_0t_1}e^{iH_0t_2}(V-H_1(0))
e^{-iH_0t_2}. \nonumber \\
   \label{order 2}
\end{eqnarray}
Since the $t_2$-integration of the second term of  Eq.(\ref{order 2}) is rewritten 
in terms of $U_1$ as 
\begin{equation}
\int_0^t dt_2\,
e^{iH_0(0)t_1}(V-H_1(0))e^{-iH_0(0)t_1} U_1(0,t_1),
\end{equation}  
it does not give a secular term {if $U_1(0,t)$ is 
secular-free\footnote{ This corresponds
to the fact that secular terms proportional to $(\lambda t)^2$ are automatically cancelled once $U_1$ 
becomes secular-free.}.}
Hence only the $t_1$-integration will give a secular term.
{ By using  $I_n$  (\ref{I-definition}) defined in the Appendix  \ref{Useful} and 
their recursion relation (\ref{I_n-recursion}),}
the second term in (\ref{order 2}) can be written, by performing the integrations,  as 
\begin{eqnarray}
\sum_{a,b\ne0}V_bV_aI_2(\omega_b,\omega_a;t) 
&=& \sum_{a,b\ne0}\frac1{i\omega_a}\delta_{a+b}V_bV_aI_1(0;t)-\sum_{a,b\ne0}\frac1{i\omega_a}V_bV_aI_1(\omega_b;t)\nonumber \\
&+&\sum_{a,b\ne0}\frac1{i\omega_a}(1-\delta_{a+b})V_bV_aI_1(\omega_b+\omega_a;t), 
\label{H2integral-free}
\end{eqnarray}
where 
\begin{equation}
\delta_{a+b}\ =\ \left\{
\begin{array}{cc}
1, & (\omega_a+\omega_b=0) \\ [2mm] 
0. & ({\rm otherwise})
\end{array}
\right.
\end{equation}
Since $I_1(0;t)=t$, the first term gives secular terms, while the other terms are secular-free. 
To eliminate the secular terms, we choose 
\begin{equation}
H_2(0)\ =\ -\sum_{a,b\ne0}\frac{1}{\omega_a}\delta_{a+b}V_bV_a. 
   \label{H_2 1st}
\end{equation}
Note that this commutes with $H_0$. 

%

\subsection{Recurrence formula for $U_n(t)$ and $H_n$ \label{sec:all-orders} }
Now we give a general prescription to remove secular terms to all orders of perturbation.
Generalizations of the explicit evaluations of multiple $t$-integrations in $U_n(t)$
are straightforward but calculations become exponentially complicated. 
(In Appendix \ref{H3integration} and \ref{4th}, 
we show the calculations for $U_3(t)$ and $U_4(t)$. )

Instead we  give an alternative method to obtain $U_n(t)$ and $H_n$ which does not
require  complicated multiple $t$-integtations.  
In order for this, we first show that secular terms proportional to 
$t^m$ ($m >1)$  are always automatically cancelled once secular terms 
in lower orders of perturbation (smaller $n$) are removed. 
Then we will give a systematic and much simpler recurrence formula
to determine $U_n(t)$ and $H_n(0)$. 

Absence of  secular terms  proportional to $t^m$ ($m>1$) is easily shown.
The $n$-th order term of the evolution operator  $U_n(t)\ :=U_n(0,t)$ defined in (\ref{xi-expansion})
satisfies the first order differential equation
\begin{eqnarray}
\frac{d}{dt}U_N(t) &=& \sum_{n=0}^{N-1}{\cal H}_{N-n}(t)U_n(t). \hspace{1cm} (N\ge1)
   \label{recursive}
\end{eqnarray}
Here  we defined $U_0(t)\ :=\ I$ and
\begin{equation}
{\cal H}_1\ :=\ V-H_1(0), \hspace{1cm} {\cal H}_n\ :=\ -i^{n-1}H_n(0). 
\end{equation}
The equation (\ref{recursive})
 can be easily guessed from the definitions of $U_n(\lambda, t)$ in (\ref{improved series}),
and proved in Appendix \ref{Proof1}.
It (\ref{recursive}) can be solved iteratively with the initial conditions $U_{n}(0)=0$ for $n\ge1$. 
Since the right-hand-side of the first order differential equation (\ref{recursive}) 
depends only on $U_n$ ($n<N$), 
$U_N(t)$ generates secular terms which are at most 
proportional to a single-power of $t$, provided that all of $U_n(t)$ for { $n < N$} are secular-free.
The following theorem and its proof in Appendix \ref{Proof-theorem}
shows that such secular terms  can be removed 
if ${\cal H}_n$ are appropriately chosen. For this, the commutativity condition (\ref{commutativity})
plays an important role.

\vspace{5mm}

Now we solve the differential equation (\ref{recursive}).
We start with $U_1(t)$ which satisfies 
\begin{equation}
\frac d{dt}U_1(t)\ =\ {\cal H}_1(t)\ =\ e^{iH_0t}(V-H_1(0))e^{-iH_0t}. 
\end{equation}
As before, we choose 
\begin{equation}
H_1(0)\ =\ V_0. 
   \label{all-H1}
\end{equation}
It is easy to integrate this equation explicitly. 
One finds 
\begin{eqnarray}
U_1(t) 
&=& \int_0^tdt_1\,\sum_{a\ne0}e^{i\omega_a t_1}V_a 
\ =\  \sum_{a\ne0}\frac1{i\omega_a}e^{i\omega_at}V_a-\sum_{a\ne0}\frac1{i\omega_a}V_a. 
\end{eqnarray}
This is secular-free. 
The result can be written as 
\begin{equation}
U_1(t)\ =\ {\cal O}_1(t)-{\cal O}_1, 
   \label{all-U1}
\end{equation}
where 
\begin{equation}
{\cal O}_1\ :=\ \sum_{a\ne0}\frac1{i\omega_a}V_a, \hspace{1cm} {\cal O}_1(t)\ :=\ e^{iH_0t}{\cal O}_1e^{-iH_0t}. 
   \label{O_1}
\end{equation}
Note that ${\cal O}_1(t)$ is a { sum of oscillating terms with frequency $\omega_a (\neq 0)$ while
${\cal O}_1$ is time-independent.}

{ To solve (\ref{recursive}) for higher $U_n(t)$ and to obtain $H_n(0)$ for $n\ge2$, }
let us introduce the following notation: 
Consider a time-independent operator ${\cal O}$. 
This can be decomposed as 
\begin{equation}
{\cal O}\ =\ [{\cal O}]+[{\cal O}]_0, \hspace{1cm} [H_0,[{\cal O}]_0]\ =\ 0. 
   \label{decompose O}
\end{equation}
It corresponds to the decomposition in (\ref{decomposition}), where $V_0=[V]_0$.
Let ${\cal O}(t)$ denote the operator $e^{iH_0t}{\cal O}e^{-iH_0t}$. 
According to (\ref{decompose O}), this can be decomposed as 
\begin{equation}
{\cal O}(t)\ =\ [{\cal O}](t)+[{\cal O}]_0. 
\end{equation}
Note that $[{\cal O}](t)$ does not contain time-independent terms since $[{\cal O}]$ does not have terms which commute with $H_0$. 

\vspace{5mm}

The following theorem holds, which can determine ${\cal H}_n$ and $U_n(t)$ iteratively. 

\begin{remark}
{\it 
Let the operators ${\cal H}_n$ and ${\cal O}_n(t)$ for $n\ge2$ be defined iteratively as 
\begin{eqnarray}
{\cal H}_n &=& -[{\cal H}_1{\cal O}_{n-1}]_0, 
   \label{def H_N} \\
\frac{d}{dt}{\cal O}_n(t) &=& \sum_{k=1}^{n-1}\left[ {\cal H}_{n-k}{\cal O}_k \right](t), 
   \label{equation O} 
\end{eqnarray}
with the initial conditions for $H_1(0)$ and ${\cal O}_1(t)$ given in (\ref{all-H1}) and (\ref{all-U1}), respectively. 
In solving (\ref{equation O}), ${\cal O}_n(t)$ is chosen to satisfy 
\begin{equation}
{\cal O}_n(t) = \left[ {\cal O}_n \right](t) .
   \label{condition O}
\end{equation}
Namely, $[{\cal O}_n]_0=0$. 
Then, the following operators $U_n(t)$ for $n\ge1$ iteratively given by 
\begin{equation}
U_n(t)\ =\ {\cal O}_n(t)-\sum_{k=1}^{n}U_{n-k}(t){\cal O}_k. 
   \label{recursion U}
\end{equation}
are secular-free, and satisfy the desired equation (\ref{recursive}).
}
\end{remark}

\begin{proof}
The proof of the theorem 
is given in Appendix \ref{Proof-theorem} where we prove that 
$U_n(t)$ iteratively defined in the theorem are {secular-free }
and satisfy the differential equation (\ref{recursive}). 
This is our main result of the present paper, and gives an iterative definition of the 
secular-free evolution equation of $U_n(t)$ and ${\cal H}_n$, which does not require
complicated multiple $t$-integrations. 
The choice of (\ref{def H_N}) 
shows that, for all $n\ge1$, $H_n(0)$ commute with $H_0$.  
The operators $U_n(t)$ defined by (\ref{recursion U}) satisfy $U_n(0)=0$, as it should be the case. 
Note also that the conditions (\ref{condition O}) are compatible with the equations (\ref{equation O}). 
The compatibility is shown in Appendix \ref{compatible}.
\end{proof}

The remaining task is to show that  secular terms are removed even when 
unperturbed Hamiltonian $H_0$ is replaced by the improved one $H_0(\lambda)$.
In order for this, we prove in Appendix \ref{higher orders} that 
it is sufficient to replace  $\omega_a$, e.g. in (\ref{H_2 1st}),
by an improved one.
Namely  $\omega_a$ needs to be renormalized and written in terms
of the eigenvalues of  the improved Hamiltonian $H_0(\lambda)$. 

\vspace{5mm}

{ In the rest of this subsection, in order to show the efficiency and simplicity
of solving the recursion equations 
(\ref{def H_N}), (\ref{equation O}) and (\ref{condition O}), we explicitly solve them to
determine  $H_n(0)$ up to fourth order of perturbation.}
Recall that we have 
\begin{equation}
\mathcal{H}_1\ =\ V-H_1\ =\ \sum_{a\ne0}V_a, \hspace{1cm} \mathcal{O}_1\ =\ \sum_{a\neq0}\frac1{i\omega_a}V_a. 
\end{equation}
By using the recursion relations, they determine ${\cal H}_2$ as  follows;
\begin{equation}
\mathcal{H}_2\ =\ -[\mathcal{H}_1\mathcal{O}_1]_0\ =\ i\sum_{a,b\neq0}\frac{1}{\omega_a}\delta_{a+b}V_bV_a. 
   \label{H_2,2}
\end{equation}
For obtaining ${\cal H}_3$, we first need to solve 
\begin{equation}
\frac d{dt}\mathcal{O}_2(t)\ =\ [\mathcal{H}_1\mathcal{O}_1](t)\ =\ \sum_{a,b\neq0}\frac{1-\delta_{a+b}}{i\omega_a}V_bV_ae^{i(\omega_a+\omega_b)t}. 
\end{equation}
This can be integrated easily and we obtain ${\cal O}_2(t)$ as
\begin{equation}
\mathcal{O}_2(t)\ =\ \sum_{a,b\neq0}\frac{1-\delta_{a+b}}{i\omega_a\cdot i(\omega_a+\omega_b)}V_bV_ae^{i(\omega_a+\omega_b)t}. 
\end{equation}
The integration constant is set to zero for ensuring the condition (\ref{condition O}). 
Then, ${\cal H}_3$ is determined as 
\begin{equation}
\mathcal{H}_3\ =\ -[\mathcal{H}_1\mathcal{O}_2]_0\ =\ \sum_{a,b,c\neq0}\frac{ 1-\delta_{a+b}}{\omega_a(\omega_a+\omega_b)}\delta_{a+b+c}V_cV_bV_a. 
   \label{H_3,2}
\end{equation}
The result coincides with the results in Appendix \ref{H3integration}. 

%

It is straightforward to proceed to ${\cal H}_4$. 
For this, we first need to determine ${\cal O}_3(t)$ by solving
\begin{eqnarray}
\frac d{dt}\mathcal{O}_3(t) 
&=& [\mathcal{H}_1\mathcal{O}_2](t)+[\mathcal{H}_2\mathcal{O}_1](t) \nonumber \\
&=& \sum_{a,b,c\neq0}\frac{1-\delta_{a+b}-\delta_{a+b+c}}{i\omega_a\cdot i(\omega_a+\omega_b)}V_cV_bV_ae^{i(\omega_a+\omega_b+\omega_c)t}+\sum_{a\neq0}\frac{1}{i\omega_a}\mathcal{H}_2V_ae^{i\omega_at}. 
\nonumber \\
\end{eqnarray}
After integration, we obtain 
\begin{eqnarray}
\mathcal{O}_3(t)&=&\sum_{a,b,c\neq0}\frac{1-\delta_{a+b}-\delta_{a+b+c}}{i\omega_a\cdot i(\omega_a+\omega_b)\cdot i(\omega_a+\omega_b+\omega_c)}V_cV_bV_ae^{i(\omega_a+\omega_b+\omega_c)t}\nonumber \\
&+&\sum_{a\neq0}\frac{1}{(i\omega_a)^2}\mathcal{H}_2V_ae^{i\omega_at}
\end{eqnarray}
Finally, ${\cal H}_4$ is determined to be 
\begin{eqnarray}
\mathcal{H}_4 
&=& -[\mathcal{H}_1\mathcal{O}_3]_0 \nonumber \\
&=& -i\sum_{a,b,c,d\neq0}\frac{1-\delta_{a+b}-\delta_{a+b+c}}{\omega_a(\omega_a+\omega_b)(\omega_a+\omega_b+\omega_c)}\delta_{a+b+c+d}V_dV_cV_bV_a\nonumber \\
&+&\sum_{a,b\neq0}\frac{1}{\omega_a^2}\delta_{a+b}V_b\mathcal{H}_2V_a. 
\label{H_4,2}
\end{eqnarray}
This expression matches with the result in Appendix \ref{4th} that is obtained by 
explicit evaluations of multiple integrations. 
 
\vspace{5mm}

We have determined $H_n(0)$ for $n=1,\cdots,4$. 
To obtain $H_0(\lambda)$, we have to promote $H_n(0)$ to $H_n(\lambda)$. 
Note that, for this purpose, we only need to know the eigenvalues of $H_0(\lambda)$ up to order ${\cal O}(\lambda^2)$. 
By substituting the eigenvalues including corrections into $H_n(0)$, we obtain an expression for $H_0(\lambda)$ valid up to order ${\cal O}(\lambda^4)$.

%


\section{Energy eigenvalues  and decay rate} \label{energy}
We have shown that secular terms can be completely removed by renormalizing the
unperturbative Hamiltonian $H_0$ into $H_0(\lambda)$ and the recurrence equation
 (\ref{def H_N})(\ref{equation O})(\ref{condition O}) 
to obtain $H_n(\lambda)$ is given. As a byproduct, this procedure provides
an  efficient algorithm for calculating perturbatively the energy eigenvalues of the full Hamiltonian $H$ 
and the decay rate of a specific state. 

\vspace{5mm}
\subsection{Energy eigenvalues}
Let us recall the explicit formula for $H_0(\lambda)$ up to order ${\cal O}(\lambda^2)$. 
For simplicity, we consider the case in which $H_0$ is non-degenerate. 
Then, $H_0(\lambda)$ is written as 
\begin{equation}
H_0(\lambda)\ =\ \sum_n|n\rangle\left[ E_n+\lambda\langle n|V|n\rangle+\lambda^2\sum_{m\ne n}\frac{\langle n|V|m\rangle\langle m|V|n\rangle}{E_n-E_m} \right]\langle n|+{\cal O}(\lambda^3). 
   \label{H_0(lambda) explicit}
\end{equation}
$H_0(\lambda)$ is diagonalized for the basis $|n\rangle$, and their eigenvalues coincide with those of the full Hamiltonian $H$ up to order ${\cal O}(\lambda^2)$. 

In fact, this is not a coincidence. 
Let ${\cal E}_n(\lambda)$ be the eigenvalues of $H$. 
The time evolution of any operator is given in terms of $e^{i{\cal E}_n(\lambda)t}$. 
 Recall that secular terms appear when the same time evolution is written in terms of $e^{iE_nt}$ which corresponds to expanding the factors $e^{i{\cal E}_n(\lambda)t}$ with respect to $\lambda$. 
Therefore, the resummation of secular terms, which is done in the improved perturbation theory, should be related to including all corrections to $E_n$. 

This can also be shown in the following manner. 
A matrix element of the operator $U(\lambda,t)$ can be written as 
\begin{eqnarray}
\langle m|U(\lambda,t)|n\rangle 
&=& \langle m|e^{iH_0(\lambda)t}e^{-iHt}|n\rangle \nonumber \\ [2mm]
&=& e^{iE_m(\lambda)t}\sum_{l}\langle m|l\rangle\hspace*{-0.5mm}\rangle\langle\hspace{-0.5mm}\langle l|n\rangle e^{-{\cal E}_n(\lambda) t}, 
   \label{U(lambda,t) expanded}
\end{eqnarray}
where 
\begin{equation}
H_0(\lambda)|n\rangle\ =\ E_n(\lambda)|n\rangle, 
\hspace{1cm} H|n\rangle\hspace{-0.5mm}\rangle\ =\ {\cal E}_n(\lambda)|n\rangle\hspace{-0.5mm}\rangle. 
\label{HandH0lambda}
\end{equation}
On the other hand, using the improved perturbation theory, one can show that the same matrix element can be written schematically as 
\begin{equation}
\langle m|U(\lambda,t)|n\rangle\ =\ e^{iE_m(\lambda)t}\sum_lc^l_{mn}e^{-iE_n(\lambda)t}. 
\end{equation}
These two expressions { show} that the following equalities
\begin{equation}
\sum_{l}\langle m|l\rangle\hspace*{-0.5mm}\rangle\langle\hspace{-0.5mm}\langle l|n\rangle e^{-{\cal E}_n(\lambda) t}\ =\ \sum_lc^l_{mn}e^{-iE_n(\lambda)t} 
\end{equation}
hold as functions of $t$ for any $m$ and $n$. 
This strongly suggests that $E_n(\lambda)={\cal E}_n(\lambda)$ holds to all orders in $\lambda$.

The fact that $H_0(\lambda)$ and $H$ share the same set of eigenvalues then implies that, as a by-product, we have now an efficient procedure for the perturbative calculation of the energy eigenvalues. 
This procedure uses the recursion relations (\ref{def H_N})(\ref{equation O})(\ref{condition O}) to obtain higher order corrections $H_n(0)$ from which $H_0(\lambda)$ is constructed. 
This works well especially when the unperturbed Hamiltonian $H_0$ does not have degeneracies. 
Since this expression is already diagonalized, the diagonal elements of $H_0(\lambda)$ give
 the correct energy eigenvalues of $H$. 
To show the efficiency of the algorithm for obtaining the energy eigenvalues, 
we study, as a simple example, an anharmonic oscillator in Appendix \ref{example}. 
 
\vspace{5mm}
\subsection{ Decay rate} 
The decaying amplitude can be obtained from the imaginary part of $H_0(\lambda)$
\begin{equation}
e^{-i H_0(\lambda) t} \rightarrow e^{-\Gamma t/2}
\end{equation}
where
\begin{equation}
\Gamma = - 2 \sum_n \lambda^n \Im (H_n(\lambda)):= \sum_n \Gamma_n .
\end{equation}
The decay rate $\Gamma$  appears as secular terms  of time-evolution
in perturbative calculations. 
They are obtained in our scheme by replacing
 one of the propagators such as  $1/\omega_a$ in  $H_n(\lambda)$ by 
\begin{equation}
\frac{1}{\omega_a - i\epsilon} = \frac{P}{\omega_a}+\pi i \delta(\omega_a),
\end{equation}
which corresponds to taking one of the intermediate states on-shell\footnote{Physically speaking, the
mathematical trick is justified when the spectrum becomes dense and 
periods of oscillations between quantum mechanical states 
 (proportional to the inverse of the energy difference) become infinitely large.} 
 and nothing but the quantum mechanical
version of the Cutkosky (cutting) rule \cite{Peskin}.

In order to reproduce the ordinary Fermi's Golden rule, it is sufficient to consider up to the second order
of perturbations for $U_n(t)$. $H_1$ does not give any imaginary part.
From ${\cal H}_2=-i H_2$ in (\ref{H_2,2}), we have 
\begin{equation}
 \Gamma_2 = 2 \pi \lambda^2 |V_0|^2
= {2 \pi}  |\sum_{\alpha,\beta} \langle  \beta|\lambda V|\alpha \rangle |^2 \delta(E_\alpha-E_\beta) 
|\alpha \rangle \langle \alpha| .
\end{equation}
Here $\Gamma_2$ gives the decay rate (Fermi's Golden rule) of each state $|\alpha \rangle$.

The next order calculation for the Fermi Golden rule can be obtained by considering $U_n(t)$ up to the
fourth order. 
From ${\cal H}_3=H_3$ in (\ref{H_3,2}), we have
\begin{equation}
  \Gamma_3=  2 \pi \lambda^3 \sum_b \left( \frac{|V_b|^2 V_0 }{\omega_b} + \frac{V_0 |V_b|^2}{\omega_b} \right) .
\end{equation}
These two terms are obtained  from the imaginary parts of $1/\omega_a$ and $1/(\omega_a+\omega_b)$.
This gives an interference term between the first and the second order perturbation for the decay amplitude.
Similarly, from ${\cal H}_4=i H_4$ in (\ref{H_4,2}), we have
\begin{equation}
  \Gamma_4= 2 \pi \lambda^4 \left( \sum_{b} \frac{|V_b|^2}{\omega_b} \right)^2
\end{equation}
Here we neglected the terms with $\delta(\omega_a+\omega_b+\omega_c)$  and $\delta(\omega_a)$
since they do not contribute to the second order generalization of Fermi's Golden rule.
It is also true for the imaginary term coming from the second term of (\ref{H_4,2}) since 
${\cal H}_2$ is a renormalization of a interaction vertex $V$ and should be treated as a whole.
Combining them, we have the decay rate of a state $|\alpha \rangle$ by 
\begin{eqnarray}
\sum_{n=1}^{4} \Gamma_n &=& 2 \pi \left| \lambda V_0 + \lambda^2  \sum_b \frac{ |V_b|^2}{\omega_b} \right|^2 
\nonumber \\
&=& 2  \pi \sum_\alpha
\left| \sum_\beta \langle \beta|\lambda V| \alpha  \rangle { +} 
\frac{ \sum_{\beta,\gamma} \langle    \beta | \lambda V| \gamma \rangle \langle \gamma |\lambda V |\alpha \rangle}
{ E_\alpha - E_\gamma  } \right|^2  
\delta(E_\alpha -E_\beta) |\alpha \rangle \langle \alpha| .
\nonumber \\
\end{eqnarray}
It is a generalized Fermi's Golden rule up to the second order of perturbation \cite{Golden}.
Higher orders can be similarly calculated, but an appropriate treatment of the double pole, the second term of ${\cal H}_4$, is required. We leave it for future investigations.

\section{Conclusions and Discussions} \label{discuss}

We have discussed an improvement of the perturbative calculation which eliminates all secular terms to all orders in the perturbative expansion. 
It turned out that the resummation of secular terms amounts to including all corrections, due to the perturbation $\lambda V$, to the energy eigenvalues of the unperturbed Hamiltonian $H_0$. 
Using this fact, we found a recursive algorithm for calculating higher order corrections to the energy eigenvalues,
as well as decay rates, which is efficient especially when the spectrum of $H_0$ does not have degeneracies. 

Multiple integrations over time $t$ in the original Dyson series (\ref{ordinary series})  generate various types of
secular terms. For example, the double integration over $t_1$ and $t_2$ of the $(-i \lambda)^2$ term
in  (\ref{ordinary series}) generates both of $(\lambda t)^2$ terms and $\lambda^2 t$ terms.
As we showed in the proof of Section \ref{sec:all-orders}, 
if the secular terms are removed order by order of perturbation,
the multiple integrations in the improved 
Dyson series in (\ref{improvedDyson}) generate only secular terms which are linear in $t$ such as $\lambda^n t$,
and secular terms with higher powers of $t$ never appear.
Thus we can remove these secular terms by renormalizing $H_0(\lambda)$ at each order of perturbation. 
In comparison with the resummation of large logarithms in quantum field theories, 
secular terms at $n$-th order  $\sim \lambda^n t$ in the improved Dyson series correspond to 
$n$-loop corrections  proportional to $\lambda^n \log (\Lambda^2)$.
In this sense, coefficients of the secular terms at $n$-th order $\lambda^n t$ determine 
coefficients of the beta function for {\it mass} renormalization. 
More precisely, the secular terms are removed by renormalization of energy eigenvalue at each level.
Thus the beta function is an operator rather than a mass parameter. 

The determination of $H_0(\lambda)$ is performed by imposing 
the commutativity condition (\ref{commutativity}).
This means that the eigenfunctions of $H_0(\lambda)$ are the same as those of the original Hamiltonian $H_0$,
although their energy eigenvalues are changed. 
The property provides a new method to calculate energy eigenvalues to all orders of perturbation
without taking care of eigenfunctions. 
Using  the notations (\ref{HandH0lambda}) and putting ${\cal E}_n(\lambda)=E_n(\lambda)$,  
the total Hamiltonian  $H$ and the improved free Hamiltonian $H_0(\lambda)$ are written as
$H=\sum {\cal E}_n(\lambda) | n \rangle\hspace*{-0.5mm}\rangle\langle\hspace{-0.5mm}\langle n|$
and $H_0(\lambda) = \sum {\cal E}_n(\lambda) | n \rangle\langle n| $. 
Thus they are  unitary equivalent:  $H_0(\lambda) = V H V^\dagger$ where $V$ is 
 a unitary operator and the evolution operator is written as
$U(\lambda, t)= e^{iH_0(\lambda)t}e^{-iHt} = V e^{iHt} V^\dagger e^{-iHt}$. 
Of course, they are unitary inequivalent with the original free Hamiltonian 
$H_0=  \sum E_n  | n \rangle\langle n| $. 
Speliotopoulos  proposed a new method to obtain energy eigenvalues
of one-dimensional anharmonic oscillators \cite{Speliotopoulos} and
then Fernandez, based on the above work, developed a method to remove secular terms 
\cite{Fernandez}. It is interesting to see how it is related to the method based on Dyson series.

We have restricted our attention to quantum mechanical systems. 
An extension to quantum field theories  would be straightforward (see e.g. \cite{deVega}).
A subtle point  is that the energy eigenstates are
no longer discrete and  we also need to renormalize various parameters
associated with UV divergences appropriately. 
It would be also interesting to apply our analysis to the quantum field theories on de Sitter spacetime. 
In this case, since the number of physical modes inside the Hubble horizon changes with time, UV and IR 
divergences are mixed up and renormalized physical parameters 
may change with time \cite{Kitazawa}. 
It is another interesting example of secular perturbations.

\vspace{1cm}

{\Large \bf Acknowledgements}

\vspace{3mm}

This work is supported in part by Grants-in-Aid for Scientific
Research (No.\ 16K05329) from 
the Japan Society for the Promotion of Science.

%
\appendix

%
\section{Useful recursion formula} \label{Useful}
{ In order to perform  multiple $t$-integrations in evaluating $U_n(0,t)$,}
 we introduce the following functions: 
\begin{equation}
I_n(\omega_1,\cdots,\omega_n;t)\ :=\ \int_0^tdt_1\cdots\int_0^{t_{n-1}}dt_n\,e^{i\omega_1t_1}\cdots e^{i\omega_nt_n}. 
\label{I-definition}
\end{equation}
If $\omega_n=0$,  this function generates a secular term { associated with $t_n$-integration.}
Otherwise, they satisfy the following recursion relations: 
\begin{eqnarray}
I_n(\omega_1,\cdots,\omega_n;t)&:=&\frac1{i\omega_n}I_{n-1}(\omega_1,\cdots,\omega_{n-2},\omega_{n-1}+\omega_n;t)\nonumber \\
&-&\frac1{i\omega_n}I_{n-1}(\omega_1,\cdots,\omega_{n-2},\omega_{n-1}). 
\label{I_n-recursion}
\end{eqnarray}
These relations enable one to determine the functions $I_n(\omega_1,\cdots,\omega_n;t)$ recursively. 
Note that $I_1(\omega,t)$ is given by
\begin{eqnarray}
& I_1(\omega,t)& =\frac{e^{i \omega t}-1}{i \omega},  \hspace{10mm} \omega \neq 0
\nonumber \\
&   I_1(0,t)& = t.
\end{eqnarray} 

\section{Explicit calculation of  $H_3(0)$} \label{H3integration}
In this appendix, we evaluate $H_3(0)$ by explicitly performing multiple integration.
$U_3(0,t)$ can be written as 
\begin{eqnarray}
U_3(0,t) 
&=& \int_0^tdt_1\,e^{iH_0t_1}H_3(0)e^{-iH_0t_1} \nonumber \\
& & +\int_0^tdt_1\int_0^{t_1}dt_2\,e^{iH_0t_1}(-iH_2(0))e^{-iH_0t_1}e^{iH_0t_2}(V-H_1(0))e^{-iH_0t_2} \nonumber \\
& & +\int_0^tdt_1\int_0^{t_1}dt_2\,e^{iH_0t_1}(V-H_1(0))e^{-iH_0t_1}e^{iH_0t_2}(-iH_2(0))e^{-iH_0t_2} \nonumber \\
& & +\int_0^tdt_1\int_0^{t_1}dt_2\int_0^{t_2}dt_3\,e^{iH_0t_1}(V-H_1(0))e^{-iH_0t_1}\nonumber \\
& & \hspace*{3cm}\times e^{iH_0t_2}(V-H_1(0))e^{-iH_0t_2}e^{iH_0t_3}(V-H_1(0))e^{-iH_0t_3}.\nonumber \\ 
   \label{3rd}
\end{eqnarray}
It turns out that the cancellation of secular terms in $U_3(0,t)$ is rather non-trivial, even though only terms proportional to $t$ appear. 

Let us first consider the second term. 
From (\ref{U3secular1}), all the secular terms
come from the $t_1$-integration and are proportional to a single power of $t$. 
It can be written as 
\begin{equation}
-i\sum_{a\ne0}H_2(0)V_aI_2(0,\omega_a;t)\ =\ -i\sum_{a\ne0}\frac1{i\omega_a}H_2(0)V_aI_1(\omega_a;t)
+i\sum_{a\ne0}\frac1{i\omega_a}H_2(0)V_aI_1(0;t). 
   \label{3rd-2}
\end{equation}
The second term produces secular terms since $I_1(0;t) =t$. 
Since $V_a \ (a \neq 0) $ do not commute with $H_0$, these secular terms cannot be
cancelled by $H_3(\lambda)$ that is required to satisfy the commutativity condition (\ref{commutativity}). 
Thus it must be cancelled by other secular terms. 

Next, the third term in (\ref{3rd}) also gives a similar kind of secular terms: 
\begin{equation}
-i\sum_{a\ne0}V_aH_2(0)I_2(\omega_a,0;t). 
   \label{3rd-3}
\end{equation}

Fortunately, both of these secular terms, (\ref{3rd-2}) and (\ref{3rd-3}), 
 are canceled by the forth term in (\ref{3rd}) which can be written as 
\begin{eqnarray}
&& \sum_{a,b,c\ne0}V_cV_bV_aI_3(\omega_c,\omega_b,\omega_a;t)  \nonumber \\
&=& \sum_{a,b,c\ne0}\frac1{i\omega_a}\delta_{a+b}V_cV_bV_aI_2(\omega_c,0;t)
-\sum_{a,b,c\ne0}\frac1{i\omega_a}\frac1{i\omega_b}\delta_{b+c}V_cV_bV_aI_1(0;t) \nonumber \\
& & +\sum_{a,b,c\ne0}\frac1{i(\omega_b+\omega_a)}\frac1{i\omega_a}\delta_{a+b+c}V_cV_bV_aI_1(0;t)+\cdots, 
\label{3rd-4}
\end{eqnarray}
where $\cdots$ indicates secular-free terms. 
The first term of (\ref{3rd-4}) 
cancels the secular terms in (\ref{3rd-3}), and the second one cancels the second term of (\ref{3rd-2}). 
The remaining secular terms, the third terms in (\ref{3rd-4}), can be canceled by choosing $H_3(0)$ as 
\begin{equation}
H_3(0)\ =\ \sum_{a,b,c\ne0}\frac{ 1- \delta_{a+b}}{\omega_a(\omega_a+\omega_b)}\delta_{a+b+c}V_cV_bV_a. 
\label{H3-0th}
\end{equation}
This commutes with $H_0$.

{ What we have found in the analysis up to the third order of perturbation is the following. 
$U_3(0,t)$ in (\ref{3rd}) is a sum of terms that are at most triple integrations over time $(t_1, t_2, t_3)$.
Thus we would have encountered a secular term that is proportional to $t^3$ or $t^2$.
However, we explicitly saw that miraculous cancellations 
remove both of $t^2$ and $t^3$ secular terms, and the only secular terms are linearly proportional to $t$.
This is the reason why we could  remove the secular term by improving the unperturbed Hamiltonian
with an appropriate choice of  $H_3(0).$
}

\section{Explicit calculation of $H_4(0)$} \label{4th}
In this appendix, we evaluate $H_4(0)$ by explicitly performing multiple integration.
Using the notations used in Section \ref{sec:all-orders}, $U_4(0,t)$ is written as a sum 
of the following terms;
\begin{eqnarray}
&&\int^t_0dt_1\ e^{iH_0t_1}\mathcal{H}_4e^{-iH_0t_1}\nonumber \\
&+&\int^t_0dt_1\int^{t_1}_0dt_2\ e^{iH_0t_1}\mathcal{H}_3e^{-iH_0t_1}e^{iH_0t_2}\mathcal{H}_1e^{-iH_0t_2}\nonumber \\
&+&\int^t_0dt_1\int^{t_1}_0dt_2\ e^{iH_0t_1}\mathcal{H}_1e^{-iH_0t_1}e^{iH_0t_2}\mathcal{H}_3e^{-iH_0t_2}\nonumber \\
&+&\int^t_0dt_1\int^{t_1}_0dt_2\ e^{iH_0t_1}\mathcal{H}_2e^{-iH_0t_1}e^{iH_0t_2}\mathcal{H}_2e^{-iH_0t_2}\nonumber \\
&+&\int^t_0dt_1\int^{t_1}_0dt_2\int^{t_2}_0dt_3\ e^{iH_0t_1}\mathcal{H}_2e^{-iH_0t_1}e^{iH_0t_2}\mathcal{H}_1e^{-iH_0t_2}e^{iH_0t_3}\mathcal{H}_1e^{-iH_0t_3}\nonumber \\
&+&\int^t_0dt_1\int^{t_1}_0dt_2\int^{t_2}_0dt_3\ e^{iH_0t_1}\mathcal{H}_1e^{-iH_0t_1}e^{iH_0t_2}\mathcal{H}_2e^{-iH_0t_2}e^{iH_0t_3}\mathcal{H}_1e^{-iH_0t_3}\nonumber \\
&+&\int^t_0dt_1\int^{t_1}_0dt_2\int^{t_2}_0dt_3\ e^{iH_0t_1}\mathcal{H}_1e^{-iH_0t_1}e^{iH_0t_2}\mathcal{H}_1e^{-iH_0t_2}e^{iH_0t_3}\mathcal{H}_2e^{-iH_0t_3}\nonumber \\
&+&\int^t_0dt_1\int^{t_1}_0dt_2\int^{t_2}_0dt_3\int^{t_3}_0dt_4\ e^{iH_0t_1}\mathcal{H}_1e^{-iH_0t_1}e^{iH_0t_2}\mathcal{H}_1e^{-iH_0t_2}\nonumber \\
&&\hspace{5cm}\times e^{iH_0t_3}\mathcal{H}_1e^{-iH_0t_3}e^{iH_0t_4}\mathcal{H}_1e^{-iH_0t_4} \nonumber \\
&=& \sum_{n=0}^3 \int_0^{t} dt_1  {\cal H}_{4-n}(t)  U_{n}(t) .
\end{eqnarray}
The operator $H_4(0)$ is determined by the requirement that the above do not generate secular terms.
The last equality is easily checked by recombining various terms into a form of $U_1(t)$, $U_2(t)$ and $U_3(t)$.
The last expression shows that the only possible secular terms are proportional to a single power of $t$ if
$U_n(t)$ for $n \le 3$ does not generate secular terms. 

Picking up only secular terms, each of these terms can be evaluated as follows;
\begin{equation}
\int^t_0dt_1\int^{t_1}_0dt_2\ e^{iH_0t_1}\mathcal{H}_3e^{-iH_0t_1}e^{iH_0t_2}\mathcal{H}_1e^{-iH_0t_2}=i\sum_{a\neq0}\frac{1}{\omega_a}\mathcal{H}_3V_at+\cdots, 
\end{equation}
\begin{equation}
\int^t_0dt_1\int^{t_1}_0dt_2\ e^{iH_0t_1}\mathcal{H}_1e^{-iH_0t_1}e^{iH_0t_2}\mathcal{H}_3e^{-iH_0t_2}=-i\sum_{a\neq0}\frac{1}{\omega_a}V_a\mathcal{H}_3te^{i\omega_at}+\cdots, 
\end{equation}
\begin{equation}
\int^t_0dt_1\int^{t_1}_0dt_2\ e^{iH_0t_1}\mathcal{H}_2e^{-iH_0t_1}e^{iH_0t_2}\mathcal{H}_2e^{-iH_0t_2}=\frac{1}{2}(\mathcal{H}_2)^2t^2,
\end{equation}
\begin{eqnarray}
&&\int^t_0dt_1\int^{t_1}_0dt_2\int^{t_2}_0dt_3\ e^{iH_0t_1}\mathcal{H}_2e^{-iH_0t_1}e^{iH_0t_2}\mathcal{H}_1e^{-iH_0t_2}e^{iH_0t_3}\mathcal{H}_1e^{-iH_0t_3}\nonumber \\
&=&\sum_{a,b\neq0}\frac{1-\delta_{a+b}}{\omega_a(\omega_a+\omega_b)}\mathcal{H}_2V_bV_at-\sum_{a,b\neq0}\frac{1}{\omega_a\omega_b}\mathcal{H}_2V_bV_at-\frac{1}{2}(\mathcal{H}_2)^2t^2+\cdots,
\end{eqnarray}
\begin{eqnarray}
&&\int^t_0dt_1\int^{t_1}_0dt_2\int^{t_2}_0dt_3\ e^{iH_0t_1}\mathcal{H}_1e^{-iH_0t_1}e^{iH_0t_2}\mathcal{H}_2e^{-iH_0t_2}e^{iH_0t_3}\mathcal{H}_1e^{-iH_0t_3}\nonumber \\
&=&\sum_{a,b\neq0}\frac{1}{\omega_a\omega_b}V_b\mathcal{H}_2V_ate^{i\omega_bt}-\sum_{a,b\neq=0}\frac{1}{\omega_a^2}\delta_{a+b}V_b\mathcal{H}_2V_at+\cdots,
\end{eqnarray}
\begin{eqnarray}
&&\int^t_0dt_1\int^{t_1}_0dt_2\int^{t_2}_0dt_3\ e^{iH_0t_1}\mathcal{H}_2e^{-iH_0t_1}e^{iH_0t_2}\mathcal{H}_1e^{-iH_0t_2}e^{iH_0t_3}\mathcal{H}_1e^{-iH_0t_3}\nonumber \\
&=&\sum_{a,b\neq0}\frac{1}{\omega_a^2}\delta_{a+b}V_bV_a\mathcal{H}_2t-\sum_{a,b\neq0}\frac{1-\delta_{a+b}}{\omega_a(\omega_a+\omega_b)}V_bV_a\mathcal{H}_2te^{i(\omega_a+\omega_b)t}-\frac{1}{2}(\mathcal{H}_2)^2t^2+\cdots,\nonumber \\
\end{eqnarray}
\begin{eqnarray}
&&\int^t_0dt_1\int^{t_1}_0dt_2\int^{t_2}_0dt_3\int^{t_3}_0dt_4\ e^{iH_0t_1}\mathcal{H}_1e^{-iH_0t_1}e^{iH_0t_2}\mathcal{H}_1e^{-iH_0t_2}\nonumber \\
&&\hspace{6cm}\times e^{iH_0t_3}\mathcal{H}_1e^{-iH_0t_3}e^{iH_0t_4}\mathcal{H}_1e^{-iH_0t_4}\nonumber \\
&=&i\sum_{a,b,c,d\neq0}\frac{1-\delta_{a+b}-\delta_{a+b+c}}{\omega_a(\omega_a+\omega_b)(\omega_a+\omega_b+\omega_c)}\delta_{a+b+c+d}V_dV_cV_bV_at\nonumber \\
&-&i\sum_{a\neq0}\frac{1}{\omega_a}\mathcal{H}_3V_at-\sum_{a,b\neq0}\frac{1-\delta_{a+b}}{\omega_a(\omega_a+\omega_b)}\mathcal{H}_2V_bV_at+\sum_{a,b\neq0}\frac{1}{\omega_a\omega_b}\mathcal{H}_2V_bV_at\nonumber \\
&+&i\sum_{a\neq0}\frac{1}{\omega_a}V_a\mathcal{H}_3te^{i\omega_at}-\sum_{a,b\neq0}\frac{1}{\omega_a\omega_b}V_b\mathcal{H}_2V_ate^{i\omega_bt}-\sum_{a,b\neq0}\frac{1}{\omega_a^2}\delta_{a+b}V_bV_a\mathcal{H}_2t\nonumber \\
&+&\sum_{a,b\neq0}\frac{1-\delta_{a+b}}{\omega_a(\omega_a+\omega_b)}V_bV_a\mathcal{H}_2te^{i(\omega_a+\omega_b)t}+\frac{1}{2}(\mathcal{H}_2)t^2+\cdots.\nonumber \\
\end{eqnarray}
Here $\cdots$ indicates terms that are free from secular terms. 
Most of the secular terms are cancelled each other. Especially secular terms proportional to $t^2$
are completely cancelled. Consequently  only the following three terms remain: 
\begin{equation}
\mathcal{H}_4t+i\sum_{a,b,c,d\neq0}\frac{1-\delta_{a+b}-\delta_{a+b+c}}{\omega_a(\omega_a+\omega_b)(\omega_a+\omega_b+\omega_c)}\delta_{a+b+c+d}V_dV_cV_bV_at-\sum_{a,b\neq0}\frac{1}{\omega_a^2}\delta_{a+b}V_b\mathcal{H}_2V_at. \nonumber
\end{equation}
Therefor, by choosing
\begin{equation}
\mathcal{H}_4=-i\sum_{a,b,c,d\neq0}\frac{1-\delta_{a+b}-\delta_{a+b+c}}{\omega_a(\omega_a+\omega_b)(\omega_a+\omega_b+\omega_c)}\delta_{a+b+c+d}V_dV_cV_bV_a+\sum_{a,b\neq0}\frac{1}{\omega_a^2}\delta_{a+b}V_b\mathcal{H}_2V_a, 
\end{equation}
all the secular terms are canceled. The choice satisfies the commutativity condition (\ref{commutativity}).

\section{Proof of (\ref{recursive})} \label{Proof1}
In order to prove that $U_n(t)\ :=U_n(0,t)$ satisfies the first order differential equation (\ref{recursive}),
we first define $\tilde{U}(\lambda,t)$ by
\begin{equation}
\tilde{U}(\lambda,t)\ :=\ \sum_{n=0}^\infty(-i\lambda)^nU_n(t), 
\end{equation}
Then one can show that this can be obtained by expanding 
\begin{equation}
\tilde{U}(\lambda,t)\ =\ e^{iH_0t}e^{-i\tilde{H}t}, \hspace{1cm} \tilde{H}\ :=\ H_0+\lambda\sum_{n=1}^\infty(-i\lambda)^{n-1}{\cal H}_n 
\end{equation}
in $\lambda$, where 
\begin{equation}
{\cal H}_1\ :=\ V-H_1(0), \hspace{1cm} {\cal H}_n\ :=\ -i^{n-1}H_n(0). 
\end{equation}
Since $\tilde{U}(\lambda,t)$ satisfies 
\begin{equation}
\frac d{dt}\tilde{U}(\lambda,t)\ =\ \sum_{n=1}^\infty(-i\lambda)^{n}{\cal H}_n(t)\cdot \tilde{U}(\lambda,t), \hspace{1cm} {\cal H}_n(t)\ :=\ e^{iH_0t}{\cal H}_ne^{-iH_0t}, 
\end{equation}
one can show that $U_n(t)$ satisfy 
\begin{eqnarray}
\frac{d}{dt}U_N(t) &=& \sum_{n=0}^{N-1}{\cal H}_{N-n}(t)U_n(t). \hspace{1cm} (N\ge1)
\end{eqnarray}


\section{Compatibility of (\ref{condition O}) and (\ref{equation O})} \label{compatible}
This can be easily seen by rewriting (\ref{equation O}) as 
\begin{equation}
i[H_0,{\cal O}_n]\ =\ \sum_{k=1}^{n-1}\left[ {\cal H}_{n-k}{\cal O}_k \right]. 
   \label{equation O-2}
\end{equation}
There exists an ${\cal O}_n$ satisfying this equation, since the right-hand side of (\ref{equation O}) is a sum of terms of the form $e^{i\omega_a t}$. 
One may assume that $[{\cal O}_n]_0=0$ since this does not contribute to the left-hand side of (\ref{equation O-2}). 



\section{Proof of the theorem}  \label{Proof-theorem}
In the following we prove that, by using the iterative definitions, (\ref{def H_N})  and (\ref{equation O}),  
the recursion relation (\ref{recursion U}) gives the secular-free solution of the equation (\ref{recursive}). 
This was already verified for $N=1$. 
For $N\ge2$, the right-hand side of (\ref{recursive}) can be written as 
\begin{eqnarray}
\sum_{n=0}^{N-1}{\cal H}_{N-n}(t)U_n(t) 
&=& {\cal H}_N(t)+\sum_{n=1}^{N-1}{\cal H}_{N-n}(t){\cal O}_n(t)-\sum_{n=1}^{N-1}{\cal H}_{N-n}(t)\sum_{k=1}^{n}U_{n-k}(t){\cal O}_k \nonumber \\
   \label{dU_N} 
\end{eqnarray}
The third term of (\ref{dU_N}) can be written as 
\begin{eqnarray}
\sum_{n=1}^{N-1}  {\cal H}_{N-n}(t)\sum_{k=1}^{n}U_{n-k}(t){\cal O}_k 
&=& \sum_{k=1}^{N-1}\sum_{n=0}^{N-k-1}{\cal H}_{N-k-n}(t)U_{n}(t){\cal O}_k  \nonumber \\
&=& \sum_{k=1}^{N-1}\frac{d}{dt}U_{N-k}(t){\cal O}_k. 
   \label{all-3rd}
\end{eqnarray}
The second term of (\ref{dU_N}) can be written as 
\begin{equation}
\sum_{n=1}^{N-1}{\cal H}_{N-n}(t){\cal O}_n(t)\ =\ {\cal H}_1(t){\cal O}_{N-1}(t)+\sum_{n=1}^{N-2}{\cal H}_{N-n}(t){\cal O}_n(t). 
   \label{2nd}
\end{equation}
One finds $[{\cal O}_n]_0=0$ since ${\cal O}_n(t)$ satisfy (\ref{condition O}), and ${\cal H}_{N-n}=[{\cal H}_{N-n}]_0$ by construction. 
These facts imply that the sum in the right-hand side of (\ref{2nd}) can be written as 
\begin{equation}
\sum_{n=1}^{N-2}{\cal H}_{N-n}(t){\cal O}_n(t)\ =\ \sum_{n=1}^{N-2}\left[ {\cal H}_{N-n}{\cal O}_n \right](t). 
\end{equation}

We decompose the first term in (\ref{2nd}) as 
\begin{eqnarray}
{\cal H}_1(t){\cal O}_{N-1}(t) 
= e^{iH_0t}{\cal H}_1{\cal O}_{N-1}e^{-iH_0t} 
= [{\cal H}_1{\cal O}_{N-1}](t)+[{\cal H}_1{\cal O}_{N-1}]_0. 
\end{eqnarray}
The second term is the only possible source of secular terms in $U_N(t)$. 
This is canceled by ${\cal H}_N(t)={\cal H}_N$ in (\ref{dU_N}). 

Taking all the above into account, the right-hand side of (\ref{dU_N}) is rewritten as
\begin{equation}
\frac d{dt}{\cal O}_N(t)-\sum_{k=1}^{N-1}\frac d{dt}U_{N-k}(t){\cal O}_k\ =\ \frac d{dt}U_N(t). 
\end{equation}
Therefore, the equation (\ref{recursive}) is satisfied for all $N\ge1$. 
Since ${\cal O}_N(t)$ is secular-free due to (\ref{condition O}), $U_N(t)$ is also secular-free. 
This completes the proof.

\section{Renormalization of energy eigenstates} \label{higher orders}
In this appendix, we show that it is necessary to
 take higher order terms in $\lambda$ into $H_n(\lambda)$ and renormalize the energy eigenvalues
 $\omega_a$ appearing in $H_n(0)$ accordingly.
We show that this can be achieved by modifying the analysis for determining $H_n(0)$ for $n\ge1$ in the previous subsection. 

As was mentioned at the end of section \ref{improved}, $U_n(\lambda,t)$ satisfy a set of differential equations. 
The explicit forms of the equations are 
\begin{equation}
\frac{d}{dt}U_N(\lambda,t)\ =\ \sum_{n=0}^{N-1}{\cal H}_{N-n}(\lambda,t)U_n(\lambda,t). \hspace{1cm} (N\ge1)
\end{equation}
where 
\begin{equation}
{\cal H}_1(\lambda)\ :=\ V-H_1(\lambda), \hspace{1cm} {\cal H}_n(\lambda)\ :=\ -i^{n-1}H_n(\lambda), 
\end{equation}
and 
\begin{equation}
{\cal H}_n(\lambda,t)\ :=\ e^{iH_0(\lambda)t}{\cal H}_n(\lambda)e^{-iH_0(\lambda)t}. 
\end{equation}
These equations have the same form with those discussed in the previous subsection. 
Therefore, $U_n(\lambda,t)$ can be constructed similarly. 

We determine higher order terms of $H_n(\lambda)$, and therefore $U_n(\lambda,t)$, by induction. 
Suppose that $H_n(\lambda)$ for $n\ge1$ have been determined up to order ${\cal O}(\lambda^{k-1})$, which satisfy the commutativity condition (\ref{commutativity}). 
This implies that $H_0(\lambda)$ has been determined up to order ${\cal O}(\lambda^k)$. 

Let $H_0^{(k)}(\lambda)$ be defined such that 
\begin{equation}
H_0^{(k)}(\lambda)\ =\ H_0(\lambda) \mbox{ mod }\lambda^{k+1}
\end{equation}
holds. 
For example, $H_0^{(1)}(\lambda)=H_0+\lambda H_1(0)$. 
Note that $H_0(\lambda)$ is formally obtained as $\lim_{k \rightarrow \infty} H_0^{(k)}(\lambda)$. 
In the following, we use the symbol $\equiv$ to means the equality modulo $\lambda^{k+1}$. 

Since $H_0^{(k)}(\lambda)$ commutes with $H_0$ by assumption, it is possible to assume that 
the eigenstates $|n\rangle$ of $H_0$ have been chosen such that they are also the eigenstates of $H_0^{(k)}(\lambda)$: 
\begin{equation}
H_0^{(k)}(\lambda)|n\rangle\ =\ E_n^{(k)}(\lambda)|n\rangle, \hspace{1cm} E_n^{(k)}(\lambda)\ 
=\ E_n+{\cal O}(\lambda). 
\end{equation}

\vspace{5mm}

Let us examine $U_1(\lambda,t)$ up to order ${\cal O}(\lambda^k)$. 
This satisfies 
\begin{equation}
\frac d{dt}U_1(\lambda,t)\ \equiv\ e^{iH_0^{(k)}(\lambda)t}(V-H_1(\lambda))e^{-iH_0^{(k)}(\lambda)t}. 
\end{equation}
In general, $V_a$ do no longer have simple commutation relations with $H_0(\lambda)$ like (\ref{decomposition}). 
To integrate this equation, we need to divide $V_a$ for $a\ne0$ further as 
\begin{equation}
V_a\ =\ \sum_{\alpha_a}V^{(k)}_{a,\alpha_a}, \hspace{1cm} [H_0^{(k)}(\lambda),V_{a,\alpha_a}^{(k)}]\ =\ \omega_{a,\alpha_a}^{(k)}(\lambda)V_{a,\alpha_a}^{(k)}. 
\label{V_decompose}
\end{equation}
where 
\begin{equation}
\omega^{(k)}_{a,\alpha_a}(\lambda)\ =\ E_m^{(k)}(\lambda)-E_n^{(k)}(\lambda)\ =\ \omega_a+{\cal O}(\lambda) 
   \label{omega_a(lambda) 1st}
\end{equation}
for some $m$ and $n$. 
Then, one obtains 
\begin{eqnarray}
U_1(\lambda,t) 
&\equiv& \sum_{a\ne0}\sum_{\alpha_a}\frac1{i\omega_{a,\alpha}^{(k)}(\lambda)}\left( e^{i\omega_{a,\alpha_a}^{(k)}(\lambda)t}-1 \right)V_{a,\alpha_a}^{(k)} \nonumber \\
& & +\int_0^tdt_1\,e^{iH_0^{(k)}(\lambda)t_1}(V_0-H_1(\lambda))e^{-iH_0^{(k)}(\lambda)t_1}. 
   \label{U_1 2nd}
\end{eqnarray}
There is no secular term in the first term of (\ref{U_1 2nd}). 
If we choose 
\begin{equation}
H_1(\lambda)\ \equiv\ V_0, 
   \label{all-H1-2}
\end{equation}
then $U_1(\lambda,t)$ is secular-free, up to order ${\cal O}(\lambda^k)$. 
Note that the commutativity condition (\ref{commutativity}) holds at this order. 

There is a difference from the case in subsection \ref{sec:all-orders}. 
At the order ${\cal O}(\lambda^k)$, the operator $e^{iH_0^{(k)}(\lambda)}V_0e^{-iH_0^{(k)}(\lambda)}$ is in general $t$-dependent. 
Therefore, naively, it does not seem to be necessary to subtract the whole $V_0$ by $H_1(\lambda)$. 
However, such $t$-dependent terms produce terms like $(e^{i{\cal O}(\lambda)t}-1)/\lambda\sim t$ which are in fact secular terms. 
Therefore, the above choice for $H_1(\lambda)$ is necessary. 

As for $U_1(0,t)$, $U_1(\lambda,t)$ can be written as 
\begin{equation}
U_1(\lambda,t)\ \equiv\ {\cal O}_1^{(k)}(\lambda,t)-{\cal O}_1^{(k)}(\lambda), 
   \label{all-U1-2}
\end{equation}
where 
\begin{equation}
{\cal O}_1^{(k)}(\lambda)\ :=\ \sum_{a\ne0}\sum_{\alpha_a}\frac1{i\omega_{a,\alpha_a}^{(k)}(\lambda)}V^{(k)}_{a,\alpha_a}. 
   \label{all-O1-2}
\end{equation}
This operator can be obtained from ${\cal O}_1$ given in (\ref{O_1}) simply by the following replacement: 
\begin{equation}
V_a \ \to\ \sum_{\alpha_a}V_{a,\alpha_a}, \hspace{1cm} \omega_a \ \to\ \omega_{a,\alpha_a^{(k)}(\lambda)}. 
   \label{replacement}
\end{equation}

Now, it is obvious that the operators $U_n(\lambda,t)$ for $n\ge2$ can be determined iteratively as 
\begin{equation}
U_n(\lambda,t)\ \equiv\ {\cal O}_n^{(k)}(\lambda,t)-\sum_{k=1}^{n}U_{n-k}(t){\cal O}_k^{(k)}(\lambda), 
\end{equation}
where ${\cal O}_n^{(k)}(\lambda)$ are determined as 
\begin{eqnarray}
{\cal H}_n(\lambda) &\equiv& -[{\cal H}_1(\lambda){\cal O}_{n-1}(\lambda)]_0, \\
\frac{d}{dt}{\cal O}_n(\lambda,t) &\equiv& \sum_{k=1}^{n-1}\left[ {\cal H}_{n-k}(\lambda){\cal O}_k(\lambda) \right](t), \\
{\cal O}_n(\lambda,t) &\equiv& \left[ {\cal O}_n(\lambda) \right](t), 
\end{eqnarray}
with the initial conditions for $H_1(\lambda)$ and ${\cal O}_1(\lambda)$ given in (\ref{all-H1-2}) and (\ref{all-O1-2}), respectively. 
Note that the notation $[{\cal O}]_0$ still indicates the part of ${\cal O}$ commuting with $H_0$, not with $H_0^{(k)}(\lambda)$. 

In this manner, one can determine $H_n(\lambda)$ for $n\ge1$ up to order ${\cal O}(\lambda^k)$. 
Then, this implies that $H_0(\lambda)$ are determined up to order ${\cal O}(\lambda^{k+1})$, and the induction proceeds. 

\vspace{5mm}

We have shown that $H_n(\lambda)$ for $n\ge1$ can be determined so that the unitary operator $U(\lambda,t)$ has no secular terms to all orders in $\lambda$. 
The iterative procedure implies that $H_n(\lambda)$ for $n\ge1$ can be obtained from $H_n(0)$ via the replacement (\ref{replacement}). 
The explicit forms of $H_n(\lambda)$ are therefore 
\begin{eqnarray}
H_1(\lambda) &=& V_0, \\
H_2(\lambda) &=& -\sum_{a,b\ne0}\sum_{\alpha_a,\beta_b}\frac1{i\omega_{a,\alpha_a}(\lambda)}\delta_{a+b}V_{b,\beta_b}V_{a,\alpha_a}, \\
H_3(\lambda) &=& \sum_{a,b,c\ne0}\sum_{\alpha_a,\beta_b,\gamma_c}\frac{1}{\omega_{a,\alpha_a}(\lambda)(\omega_{a,\alpha_a}(\lambda)+\omega_{b,\beta_b}(\lambda))}\delta_{a+b+c}V_{c,\gamma_c}V_{b,\beta_b}V_{a,\alpha_a},\nonumber \\ 
\label{lastresults}
\end{eqnarray}
and so on, where 
\begin{equation}
V_{a,\alpha_a}\ :=\ \lim_{k\to\infty}V_{a,\alpha_a}^{(k)}, \hspace{1cm} \omega_{a,\alpha_a}(\lambda)\ :=\ \lim_{k\to\infty}\omega^{(k)}_{a,\alpha_a}(\lambda). 
\end{equation}

%

\section{Anharmonic Oscillator} \label{example}
As an application of the general method explained in section \ref{energy}, we consider an anharmonic oscillator as a simple example;
\begin{eqnarray}
H&=&H_0+\lambda V\nonumber \\
&=&\frac{1}{2}p^2+\frac{\omega}{2}x^2+\lambda x^4\nonumber \\
&=&\omega(\hat{N}+\frac{1}{2})+\frac{\lambda}{4\omega^2}(\hat{a}+\hat{a}^\dagger)^4,
\end{eqnarray}
where $x=(\hat{a}+\hat{a}^\dagger)/\sqrt{2\omega} ,\ p=-i\sqrt{\omega /2}(\hat{a}-\hat{a}^\dagger),\ [\hat{a},\hat{a}^\dagger]=1$ and $\hat{N}=\hat{a}^\dagger \hat{a}.$
Eigenstates of $H_0$ are constructed as usual, $|n \rangle=(\hat{a}^\dagger)^n|0\rangle /\sqrt{n!} ,\ 
H_0|n\rangle=\omega(n+1/2)    |n\rangle$.
In this example, there are no degeneracies of $H_0$ 
 and we do not need  to diagonalize $H_0(\lambda)$ at each order in $\lambda$.
The interaction term $V=(\hat{a}+\hat{a}^\dagger)^4/4\omega^2$ can be decomposed as
\begin{eqnarray}
V=V_{-4}+V_{-2}+V_0+V_2+V_4,\hspace{1cm}[H_0,V_a]=a \omega  V_a,
\end{eqnarray}
where 
\begin{eqnarray}
&&V_4=\frac{1}{4\omega^2}(\hat{a}^\dagger)^4,\\
&&V_2=\frac{1}{4\omega^2}\left((\hat{a}^\dagger)^3\hat{a}+(\hat{a}^\dagger)^2\hat{a}\hat{a}^\dagger+\hat{a}^\dagger \hat{a}(\hat{a}^\dagger)^2+\hat{a}(\hat{a}^\dagger)^3\right),\\
&&V_0=\frac{1}{4\omega^2}\left((\hat{a}^\dagger)^2\hat{a}^2+\hat{a}^\dagger \hat{a}\hat{a}^\dagger \hat{a}+\hat{a}^\dagger \hat{a}^2\hat{a}^\dagger+\hat{a}(\hat{a}^\dagger)^2\hat{a}+\hat{a}\hat{a}^\dagger \hat{a}\hat{a}^\dagger+\hat{a}^2(\hat{a}^\dagger)^2\right),\\
&&V_{-2}=\frac{1}{4\omega^2}\left(\hat{a}^\dagger \hat{a}^3+\hat{a}\hat{a}^\dagger \hat{a}^2+\hat{a}^2\hat{a}^\dagger \hat{a}+\hat{a}^3\hat{a}^\dagger\right),\\
&&V_{-4}=\frac{1}{4\omega^2}\hat{a}^4.
\end{eqnarray}
We can simplify them by sorting creation and annihilation operators to construct $\hat{N}$.  

\vspace{5mm}

We now calculate $H_n (\lambda)$ following the general prescriptions in the previous section.
First, at the leading order ${\cal O}(\lambda)$, $H_1(\lambda)$ is given by
\begin{equation}
H_1 (\lambda)=V_0=\frac{3}{4\omega^2}(2\hat{N}^2+2\hat{N}+1) .
\end{equation}
Then we get 
\begin{equation}
H_0(\lambda)=\omega(\hat{N}+\frac{1}{2})+\frac{3\lambda}{4\omega^2}(2\hat{N}^2+2\hat{N}+1)+O(\lambda^2).
\end{equation}

At the next order, we calculate $H_2(\lambda)$ as 
 in (\ref{H_2,2}) to obtain
\begin{eqnarray}
H_2(\lambda)&=&-\left[\frac{1}{-4\omega}V_{4}V_{-4}+\frac{1}{4\omega}V_{-4}V_{4}+\frac{1}{-2\omega}V_{2}V_{-2}+\frac{1}{2\omega}V_{-2}V_{2}\right]+O(\lambda)\nonumber \\
&=&-\frac{1}{8\omega^5}(34\hat{N}^3+51\hat{N}^2+59\hat{N}+21)+O(\lambda).
\end{eqnarray}
Thus, $H_0(\lambda)$ becomes
\begin{eqnarray}
H_0(\lambda)&=&\omega(\hat{N}+\frac{1}{2})+\frac{3\lambda}{4\omega^2}(2\hat{N}^2+2\hat{N}+1)\nonumber \\
&-&\frac{\lambda^2}{8\omega^5}(34\hat{N}^3+51\hat{N}^2+59\hat{N}+21)+O(\lambda^3)
\nonumber \\
\end{eqnarray}
up to  ${\cal O}(\lambda^2)$.\\

At the third order in $\lambda$, we calculate $H_3(\lambda)$.
From (\ref{H_3,2}), $H_3 (\lambda)$ is given by
\begin{eqnarray}
H_3 (\lambda)&=&\frac{1}{-4\omega(-4\omega+2\omega)}V_{2}V_{2}V_{-4}+\frac{1}{2\omega(2\omega-4\omega)}V_{2}V_{-4}V_{2}\nonumber \\
&&+\frac{1}{2\omega(2\omega+2\omega)}V_{-4}V_{2}V_{2}+\frac{1}{4\omega(4\omega-2\omega)}V_{-2}V_{-2}V_{4}
\nonumber \\
&&+\frac{1}{-2\omega(-2\omega+4\omega)}V_{-2}V_{4}V_{-2}+\frac{1}{-2\omega(-2\omega-2\omega)}V_{4}V_{-2}V_{-2}+O(\lambda)\nonumber \\
&=&\frac{3}{16\omega^8}\left(40\hat{N}^4+80\hat{N}^3+164\hat{N}^2+124\hat{N}+42\right)+O(\lambda).
\end{eqnarray}
In order to take ${\cal O}(\lambda)$ corrections to $H_2(\lambda)$ into account, 
 we decompose each $V_a$, according to (\ref{V_decompose}), as
\begin{equation}
V_a=\sum_{n}V^{(1)}_{a,n} .
\end{equation}
Here we have
\begin{eqnarray}
&& 
V^{(1)}_{4,n}=\frac{1}{4\omega^2}\sqrt{\frac{(n+4)!}{n!}}|n+4\rangle \langle n|\\
&& 
V^{(1)}_{-4,n}=\frac{1}{4\omega^2}\sqrt{\frac{n!}{(n-4)!}}|n-4\rangle \langle n|\\
&& 
V^{(1)}_{2,n}=\frac{1}{2\omega^2}(2n+3)\sqrt{\frac{(n+2)!}{n!}}|n+2\rangle \langle n|\\
&& 
V^{(1)}_{-2,n}=\frac{1}{2\omega^2}(2n-1)\sqrt{\frac{n!}{(n-2)!}}|n-2\rangle \langle n|.
\end{eqnarray}
They  satisfy
\begin{eqnarray}
&[H^{(1)}_0(\lambda),V^{(1)}_{a,n}]=\omega^{(1)}_{a,n}(\lambda)V^{(1)}_{a,n},\hspace{1cm}\omega^{(1)}_{a,n}(\lambda)=E^{(1)}_{n+a}(\lambda)-E^{(1)}_{n}(\lambda),&\\
&\displaystyle E^{(1)}_{n}(\lambda)=\omega(n+\frac{1}{2})+\frac{3\lambda}{4\omega^2}(2n^2+2n+1)+O(\lambda^2).&
\end{eqnarray}
Then, $H_2(\lambda)$ becomes 
\begin{eqnarray}
H_2(\lambda) 
&=&-\sum_{m,n}\left[\frac{1}{\omega^{(1)}_{-4,m}(\lambda)}V^{(1)}_{4,n}V^{(1)}_{-4,m}
+\frac{1}{\omega^{(1)}_{4,m}(\lambda)}V^{(1)}_{-4,n}V^{(1)}_{4,m} \right.
\nonumber  \\ 
&& 
\left.
+\frac{1}{\omega^{(1)}_{-2,m}(\lambda)}V^{(1)}_{2,n}V^{(1)}_{-2,m}
+\frac{1}{\omega^{(1)}_{2,m}(\lambda)}V^{(1)}_{-2,n}V^{(1)}_{2,m}\right]
+O(\lambda^2). \nonumber \\
\end{eqnarray}
The first term in the right-hand side can be written as
\begin{eqnarray}
&&-\frac{1}{(4\omega^2)^2}\sum_{n,m}\left(-4\omega-\frac{3\lambda}{\omega^2}(4m-6)+O(\lambda^2)\right)^{-1}\sqrt{\frac{(n+4)!m!}{n!(m-4)!}}\delta_{n,m-4}|n+4\rangle \langle m|\nonumber \\
&=&\frac{1}{2^6\omega^5}(\hat{N}^4-6\hat{N}^3+11\hat{N}^2-6\hat{N})\nonumber \\
&-&\frac{3\lambda}{2^7\omega^8}(2\hat{N}^5-15\hat{N}^4+40\hat{N}^3-45\hat{N}^2+18\hat{N})+O(\lambda^2),\nonumber
\end{eqnarray}
where we used $m|m\rangle=\hat{N}|m\rangle$ and $\sum_m|m\rangle \langle m|=1$.
Other terms are also written in a similar form and $H_2(\lambda)$ becomes
\begin{eqnarray}
H_2(\lambda)
&= & -\frac{1}{8\omega^5}(34\hat{N}^3+51\hat{N}^2+59\hat{N}+21) \nonumber \\ 
&&+\frac{3\lambda}{16\omega^8}(85\hat{N}^4+170\hat{N}^3+308\hat{N}^2+223\hat{N}+69)+O(\lambda^2).\nonumber \\
\end{eqnarray}
Summing up these $O(\lambda^3)$ contributions, we obtain 
\begin{eqnarray}
H_0(\lambda)
&=&\omega(\hat{N}+\frac{1}{2})+\frac{3\lambda}{4\omega^2}(2\hat{N}^2+2\hat{N}+1)-\frac{\lambda^2}{8\omega^5}(34\hat{N}^3+51\hat{N}^2+59\hat{N}+21)\nonumber \\
&&+\frac{3\lambda^3}{16\omega^8}(125\hat{N}^4+250\hat{N}^3+472\hat{N}^2+347\hat{N}+111)+O(\lambda^4).
\label{H0-3}
\end{eqnarray}
 Its eigenvalues coincide with those of the full Hamiltonian $H$ obtained by an ordinary method in the Schr\"odinger picture. 
It should be emphasized that we did not need to determine the perturbative corrections to the wave functions for obtaining this result. 
We can further calculate $H_0(\lambda)$ at any order in $\lambda$ in a similar way.

\vspace{1cm}

\end{document}